\documentclass[aps,prb,amsmath,amssymb,twocolumn]{revtex4}
\usepackage{graphics}

\begin{document}

\title{Generating Function in Quantum Mechanics: An Application to Counting Problems}
\author{Li Han}\email{han-l04@mails.thu.edu.cn}
%\author{Yong Guo}
\affiliation{Department of Physics, Tsinghua University, Beijing
100084, People's Republic of China}
%\date{\today}
\begin{abstract}
In this paper we present a generating function approach to two
counting problems in elementary quantum mechanics. The first is to
find the total ways of distributing $N$ identical particles among
$m$ different states. The second is to find the degeneracies of
energy levels in a quantum system with multiple degrees of freedom.
Our approach provides an alternative to the methods in textbooks.
\end{abstract} \maketitle

\section{Introduction}

Generating function (GF) is a terminology frequently used in
combinatorial mathematics as a bridge between discrete mathematics
and continuous analysis. It functions like a clothesline on which we
hang up a sequence of numbers for display \cite{Wilf}. A GF may have
different aliases in different contexts, e.g. partition function in
statistical physics \cite{Huang} and $Z$-transform in signal
analysis \cite{Oppenheim}. It can be used to find certain
characteristics of a sequence like expectation value or standard
deviation \cite{Papoulis}. Sometimes that will be much more
efficient than a direct evaluation.

There are many ways to construct a GF for a sequence, the simplest
of which is using power series. Given $\{a_n\}$ ($n\geq0$), we may
construct its GF as
\begin{eqnarray}
F(t) = \sum_{n\geq0} a_n t^n.
\end{eqnarray}
Note that $t$ can take a continous value and $F(t)$ is analytic with
respect to $t$, which may give us some computational power using
techniques of derivation or integration.

The paper is organized as follows. Section II shows how a GF can
find the number of distinct ways of assigning $N$ identical bosons
or fermions to $m$ different states. Section III discusses the GF
approach to find the degeneracies of energy levels of a harmonic
oscillator and a confined free particle both in higher dimensions.
In our discussion, we follow the notation $[t^n]\,F(t)$ used in Ref.
\cite{Wilf} to mean the coefficient of $t^n$ term in the power
series expansion of $F(t)$. For example, we have
$[t^n]\,\mathrm{e}^t=1/{n!}$ $(n\geq0)$.

\section{finding total ways of distributing identical particles}
The problem of how many ways there are to distribute $N$ identical
particles among $m$ different states lies at the heart of quantum
statistics. It determines the specific distribution function that
identical particles obey, i.e. Bose-Einstein distribution for bosons
and Fermi-Dirac distribution for fermions. Textbooks generally
tackle the problem by a combinatorial method \cite{Griffiths}. Here
we show how to solve it by a GF.

\subsection{Identical Fermions}
Pauli's exclusion principle has that a quantum state can accommodate
one fermion at most. This makes the counting problem to be much
easier than that of bosons. The number is simply the binomial
coefficient $m$ choose $N$:
\begin{eqnarray}
{m \choose N}=\frac{m!}{N!\,(m-N)!}.
\end{eqnarray}

We can do it in another way. The occupation number for fermions is
$0$ and $1$ only. We thus associate each state with a GF
$F(t)=t^0+t^1=1+t$. It is obvious that the coefficient of term $t^N$
in the expansion of $F^m(t)$ is exactly the total ways that we can
pick out $N$ candidates from $m$ states: ($N\leq m$)
\begin{widetext}
 \vspace{-5mm}
\begin{eqnarray}
 {[t^N]} F^m (t) & = & {[t^N]}\,(1+t)^m\nonumber\\
 & = & \frac{1}{N!}\,{[t^0]}\,\frac{\mathrm{d}^N}{\mathrm{d}x^N}\,(1+t)^m\nonumber\\
 & = & \frac{1}{N!}\,{[t^0]}\,m(m-1)\cdots(m-N+1)(1+t)^{m-N}\nonumber\\
 & = & \frac{m!}{N!\,(m-N)!}\nonumber\\
 & = & {m \choose N}.
\end{eqnarray}
\end{widetext}

The GF approach recovers the binomial coefficient $m$ choose $N$.
Furthermore, the approach can be easily adapted to the boson case.

\subsection{Identical Bosons}

There can be an arbitrary number of bosons in one quantum state in
principle, i.e. the occupation number can be $0$,$1$,$2$,$\cdots$.
It is not easy to figure out a combinatorial method to find the
total ways in this case. With the help of GF, we now associate each
quantum state with $B(t)=t^0+t^1+t^2+\cdots=1+t+t^2+\cdots$. Notice
that $B(t)$ has an explicit form when $t$ is within its region of
convergence:
\begin{eqnarray}
B(t)=\frac{1}{1-t},\quad \vert t\vert<1.
\end{eqnarray}

Just as what we have done for fermions, we multiply $m$ copies of
$B(t)$ together. As a result, the number of total ways to distribute
$N$ bosons among $m$ different states is just the coefficient of
term $t^N$ in the expansion of $B^m(t)$:
\begin{widetext}
 \vspace{-5mm}
\begin{eqnarray}
 {[t^N]} B^m (t) & = & {[t^N]}\,\frac{1}{(1-t)^m}\nonumber\\
 & = & \frac{1}{N!}\,{[t^0]}\,\frac{\mathrm{d}^N}{\mathrm{d}t^N}\,\frac{1}{(1-t)^m}\nonumber\\
 & = & \frac{1}{N!}\,{[t^0]}\,(-m)(-m-1)\cdots(-m-N+1)\frac{(-1)^N}{(1-t)^{m+N}}\nonumber\\
 & = &\frac{(N+m-1)!}{N!\,(m-1)!}\nonumber\\
 & = & {N+m-1 \choose N}.
\end{eqnarray}
\end{widetext}

This is exactly the same as what is derived from a combinatorial
method \cite{Huang, Griffiths}, but we are using techniques from
calculus for continuous variables. The GF approach also has an
advantage when the occupation numbers are distributed in some random
way, e.g., $1$,$2$,$5$,$\cdots$.

\section{finding degeneracies of energy levels}

Energy levels are usually quantized in a quantum system.
Furthermore, degeneracies often occur in a system with independent
degrees of freedom. In this section we employ the GF to count the
degeneracies.

Problem of this kind can be formulated as follows. Suppose the
system Hamiltonian $H$ can be written as
\begin{eqnarray}
 H=H_{\mathrm{I}}+H_{\mathrm{II}}+H_{\mathrm{III}}+\cdots,
\end{eqnarray}
where the subscripts
$\mathrm{I}$,$\mathrm{II}$,$\mathrm{III}$,$\cdots$ represent
independent degrees of freedom. Suppose further that the energy
spectrum of $H_{\Delta}$ is $\{E_{\Delta}(n),n\}$ with degeneracies
$\{d_{\Delta}(n),n\}$,
$\Delta=\mathrm{I},\mathrm{II},\mathrm{III},\cdots$. Since the
degrees of freedom are independent, the Schr\"{o}dinger equation of
the system is separable. Therefore, the energy spectrum of the
system Hamiltonian $H$ consists of all possible sums of
$\{E_{\Delta}(n),n,\Delta\}$. The degeneracy of a certain energy
level $E$ is then given by the total number of $m$-tuples
\cite{tuple} $(n_1,n_2,\cdots,n_m)$ that satisfy
\begin{eqnarray}
 E=E_{\mathrm{I}}(n_1)+E_{\mathrm{II}}(n_2)+E_{\mathrm{III}}(n_3)+\cdots,
\end{eqnarray}
with $m$ being the total number of degrees of freedom.

Now we associate each degree of freedom $\Delta$ with a GF
\begin{eqnarray}
 G_{\Delta}(t)=\sum_n d_{\Delta}(n)\,t^{E_{\Delta}(n)}
\end{eqnarray}
and multiply them together over all $\Delta$. As a result, we obtain
a polynomial \cite{polynomial} of $t$, whose coefficients give
exactly the degeneracies we want. To be specific, the degeneracy of
energy level $E$ is given by
\begin{eqnarray}
 d(E)= {[t^E]}\,\prod_{\Delta}G_{\Delta}(t).
\end{eqnarray}

\subsection{Harmonic Oscillator}
A harmonic oscillator (HO) is not only exactly soluble in both
classical and quantum physics, but also of great physical relevance.
For example, the problem of finding the Landau levels of an electron
in a magnetic field reduces to finding energy levels of an
equivalent HO. A 1D HO has a simple structure of energy spectrum,
whose levels are non-degenerate and equally spaced:
\begin{eqnarray}
 E_\mathrm{1D}(n)=\left(n+\frac12\right)\hbar\omega,\quad n\geq0.
\end{eqnarray}

Consider now a 2D homogeneous HO with potential
\begin{eqnarray}
 V(x,y)=\frac12 m\omega^2\left(x^2+y^2\right).
\end{eqnarray}
Its energy spectrum can be solved in a planar polar coordinate
system, noting that $V(x,y)=V(\rho)=\frac12 m\omega^2 \rho^2$. A
much simpler way is to treat the 2D system as two uncoupled 1D HOs.
Consequently, the energy levels are determined by a couple of
integers $(n_x,n_y)$ as
\begin{eqnarray}
 E_\mathrm{2D}(n)=\left(n_x+n_y+1\right)\hbar\omega,\quad n\geq0,
\end{eqnarray}
where $n\equiv n_x+n_y$. For a given integer $n$, there correspond
$n+1$ different couples of quantum numbers $(n_x,n_y)$, namely,
$(0,n)$, $(1,n-1)$, $\cdots$, $(n,0)$. The degeneracy of energy
level $E_\mathrm{2D}(n)$ is therefore
\begin{eqnarray}
 d_\mathrm{2D}(n)=n+1.
\end{eqnarray}

Now we apply the GF approach to find the degeneracy. Subtracting the
zero point energy $\frac12 \hbar\omega$ from the spectrum of each 1D
HO and setting $\hbar \omega=1$, we obtain
$\bar{E}_{\mathrm{2D}}(n)=n_x+n_y$, $n_x,n_y=0,1,2,\cdots$. Then we
associate each 1D HO with a GF
\begin{eqnarray}
 H_{\mathrm{1D}}(t)=\sum_{n\geq 0}t^n=1+t+t^2+\cdots,
\end{eqnarray}
which reduces to $1/(1-t)$ for $\vert t\vert<1$. The degeneracy
$d_\mathrm{2D}(n)$ of energy level $\bar{E}_\mathrm{2D}(n)$ is
therefore given by
\begin{eqnarray}
 d_\mathrm{2D}(n) & = & {[t^n]}\,H_{\mathrm{1D}}^2(t)\nonumber\\
 & = & {[t^n]}\,\frac{1}{(1-t)^2}\nonumber\\
 & = & {n+2-1 \choose n}\nonumber\\
 & = & n+1,
\end{eqnarray}
using result from section II-B.

We can go further to find the degeneracies of energy levels for a 3D
homogeneous HO whose potential is
\begin{eqnarray}
 V(r)=\frac12 m\omega^2
 r^2=\frac12 m\omega^2\left(x^2+y^2+z^2\right).
\end{eqnarray}
It is straightforward to write down that
\begin{eqnarray}
 E_\mathrm{3D}(n) & = & \left(n_x+n_y+n_z+\frac32\right)\hbar\omega\nonumber\\
 & \equiv & \left(n+\frac32\right)\hbar\omega,\quad n\geq0,
\end{eqnarray}
and the corresponding degeneracy
\begin{eqnarray}
 d_\mathrm{3D}(n) & = & {[t^n]}\,H_{\mathrm{1D}}^3(t)\nonumber\\
 & = & {[t^n]}\,\frac{1}{(1-t)^3}\nonumber\\
 & = & {n+3-1 \choose n}\nonumber\\
 & = & \frac12 (n+1)(n+2).
\end{eqnarray}
The sequence of $d_\mathrm{3D}(n)$ goes like
$1$,$3$,$6$,$10$,$15$,$21$,$\cdots$, which is consistent with the
results derived by solving in spherical coordinates.

It is not difficult to generalize above to the case of inhomogeneous
HOs, e.g. $V(x,y)=\frac12 m(\omega_x^2\,x^2+\omega_y^2\,y^2)$. As
long as the ratio $\omega_x / \omega_y$ is some rational number, we
can always find appropriate GFs to calculate the degeneracies of the
system energy levels.

\subsection{Confined Free Particle}

\begin{table}
\begin{ruledtabular}
\begin{tabular}{cccc}
Energy intervals ($E_0$) & 1D & 2D & 3D\\
\hline

%Copy from output of Mathematica code
1-10000 & 100 & 7754 & 511776 \\
10001-20000 & 41 & 7816 & 945684 \\
20001-30000 & 32 & 7816 & 1227826 \\
30001-40000 & 27 & 7821 & 1456239 \\
40001-50000 & 23 & 7843 & 1653737 \\
50001-60000 & 21 & 7819 & 1829268 \\
60001-70000 & 20 & 7842 & 1990061 \\
70001-80000 & 18 & 7833 & 2138555 \\
80001-90000 & 18 & 7830 & 2277886 \\
90001-100000 & 16 & 7859 & 2408841 \\
100001-110000 & 15 & 7827 & 2533047 \\
110001-120000 & 15 & 7840 & 2651325 \\
120001-130000 & 14 & 7846 & 2765036 \\
130001-140000 & 14 & 7838 & 2873651 \\
140001-150000 & 13 & 7837 & 2978653 \\
150001-160000 & 13 & 7835 & 3079948 \\
160001-170000 & 12 & 7847 & 3178604 \\
170001-180000 & 12 & 7845 & 3273810 \\
180001-190000 & 11 & 7842 & 3366202 \\
190001-200000 & 12 & 7844 & 3456580 \\
\hline
Total States & 447 & 156634 & 46596729 \\
%\hline
Rough Value \footnote{Rough value of total states within energy $E$
estimated by the quasi-continuous approximation. For 1D,
$N=E^{1/2}$. For 2D, $N=(\pi/4)E$. For 3D, $N=(\pi/6)E^{3/2}$.}
& $4.47\times 10^2$ & $1.57\times 10^5$ & $4.68\times 10^7$ \\

\end{tabular}
\end{ruledtabular}
\caption{\label{Degeneracies}\quad List of number of states in 20
10000$E_0$-long energy intervals for a 1D, 2D and 3D confined free
particle, respectively. $E_0={\pi^2 \hbar^2/2mL^2}$.}
\end{table}

Consider first a 1D free particle confined to some region of length
$L$. It has non-degenerate energy levels \cite{spin} as
\begin{eqnarray}
E_{\mathrm{1D}}(n)=\frac{\pi^2 \hbar^2}{2mL^2}\,n^2,\quad n\geq 1.
\end{eqnarray}
In the following, we choose $L$ such that $E_{\mathrm{1D}}(n)=n^2$
for simplicity. Accordingly, the levels $E_{\mathrm{1D}}(n)$ have a
GF as
\begin{eqnarray}
 F_{\mathrm{1D}}(t)=\sum_{n\geq 1}t^{n^2}=t+t^4+t^9+\cdots.
\end{eqnarray}

$F_{\mathrm{1D}}(t)$ can be used to find the degeneracies of energy
levels of a free particle confined in a square of $L\times L$ or
even a cube of $L\times L\times L$. Finding the degeneracies in each
case (1D, 2D and 3D) is a basic problem in quantum mechanics. In
solid state physics \cite{Kittel}it is also concerned with the
density of states per unit energy $n(E)\equiv
\mathrm{d}N/\mathrm{d}E$. In two dimensions, the density of states
$n_{\mathrm{2D}}(E)$ is independent of the energy $E$, whereas
$n_{\mathrm{1D}}(E)\propto E^{-1/2}$ in 1D and
$n_{\mathrm{3D}}(E)\propto E^{1/2}$ in 3D. Textbooks \cite{Kittel}
usually derive the density of states by using a quasi-continuous
approximation of the distribution of energy levels in $k$-space.

\begin{figure}
\scalebox{0.4}[0.4]{\includegraphics{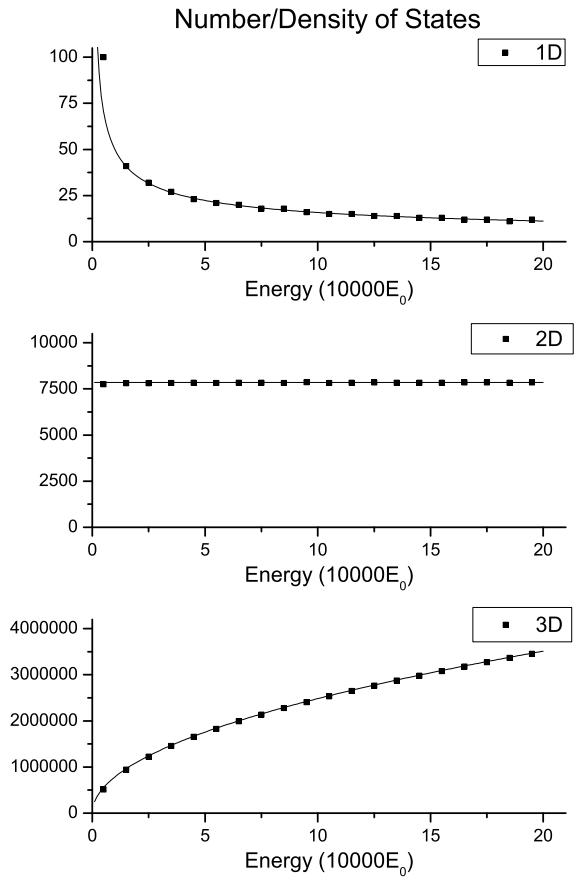}}
\caption{\label{Density Plot}\ List plots (dots) of Table
\ref{Degeneracies}, which approximate the density of states versus
energy curve (solids) $n(E)=\mathrm{d}N/\mathrm{d}E$.
$n_{\mathrm{1D}}(E)\propto E^{-1/2}$,
$n_{\mathrm{2D}}(E)=\mathrm{const}$ and $n_{\mathrm{3D}}(E)\propto
E^{1/2}$.}
\end{figure}

Below we use $F_{\mathrm{1D}}(t)$ to find the level degeneracies in
2D and 3D. Following Eq. (9), the degeneracies are
\begin{eqnarray}
 d_{\mathrm{2D}}(E) & = & [t^E]\,F_{\mathrm{1D}}^2(t),\\
 d_{\mathrm{3D}}(E) & = & [t^E]\,F_{\mathrm{1D}}^3(t).
\end{eqnarray}
Unfortunately, series like $F_{\mathrm{1D}}(t)$ does not have a
known explicit sum. Thanks to the advanced math softwares like
Mathematica, we can use a computer to perform the evaluation.
Mathematica's ``CoefficientList" function can extract the
coefficients in the expansion of Eqs. (20-22). Then we are able to
count the total number of states whose energies fall into the
interval $[E_1,E_2]$. Table \ref{Degeneracies} gives a detailed list
of number of states within certain energy intervals for a 1D, 2D and
3D free particle, respectively. The list plots of the number of
states are further presented in Fig. \ref{Density Plot}. There are
totally $20$ energy intervals and each interval has a length of
$10000$. Since the interval length is far more than the minimum
separation between energy levels, the quasi-continuous approximation
is valid and the plots will reflect approximately the dependence of
density of states on the energy. A comparison between the discrete
number of states and the continuous density of states is made in
Fig. \ref{Density Plot} (depicted as dots and solid lines,
respectively). It can be seen that the exact numbers of states
derived by a GF approach comply well with $n_{\mathrm{1D}}(E)\propto
E^{-1/2}$, $n_{\mathrm{2D}}(E)=\mathrm{const}$ and
$n_{\mathrm{3D}}(E)\propto E^{1/2}$.

\begin{acknowledgements}
This work was supported by the National Natural Science Foundation
of China (No. 10474052).
\end{acknowledgements}


\begin{references}
\bibitem{tuple} If an element $n_i$ in the tuple corresponds to a level of
degeneracy $d_i$, that tuple should be counted for $d_i$ times.
\bibitem{polynomial} Assume the energy levels are quantized so that
their separations are multiples of some unit. That is usually the
case in a simple quantum system.
\bibitem{spin} The spin degeneracies are excluded from the
discussion.
\end{references}
\end{document}